\documentclass[fleqn,10pt]{wlscirep}
\usepackage[utf8]{inputenc}
\usepackage[T1]{fontenc}
\usepackage{float}
\usepackage{caption}
\usepackage{subcaption}
\usepackage{csquotes}

\title{COVID-19 Vaccine Hesitancy on Social Media: Building a Public Twitter Dataset of Anti-vaccine Content, Vaccine Misinformation and Conspiracies}

\author[1,*,+]{Goran Muric}
\author[1,+]{Yusong Wu}
\author[1,2,3]{Emilio Ferrara}

\affil[1]{University of Southern California, Information Sciences Institute}
\affil[2]{Department of Computer Science, University of Southern California}
\affil[3]{Annenberg School for Communication and Journalism, University of Southern California}

\affil[*]{corresponding author(s): Goran Muric (gmuric@isi.edu)}
\affil[+]{These authors contributed equally to this work}

\begin{abstract}
False claims about COVID-19 vaccines can undermine public trust in ongoing vaccination campaigns, thus posing a threat to global public health. Misinformation originating from various sources has been spreading online since the beginning of the COVID-19 pandemic. In this paper, we present a dataset of Twitter posts that exhibit a strong anti-vaccine stance. The dataset consists of two parts: a) a streaming keyword-centered data collection with more than 1.8 million tweets, and b) a historical account-level collection with more than 135 million tweets. The former leverages the Twitter streaming API to follow a set of specific vaccine-related keywords starting from mid-October 2020. The latter consists of all historical tweets of 70K accounts that were engaged in the active spreading of anti-vaccine narratives. We present descriptive analyses showing the volume of activity over time, geographical distributions, topics, news sources, and inferred account political leaning. This dataset can be used in studying anti-vaccine misinformation on social media and enable a better understanding of vaccine hesitancy. In compliance with Twitter's Terms of Service, our anonymized dataset is publicly available at: \href{https://github.com/gmuric/avax-tweets-dataset}{https://github.com/gmuric/avax-tweets-dataset}

\end{abstract}
\begin{document}

\flushbottom
\maketitle
\thispagestyle{empty}
\section*{Introduction}
The opposition to vaccination dates back to the 1800s, right after the English physician Edward Jenner created the first vaccine in human history. The opponents were loud and could be found in all segments of society: religious communities protested the unnaturalness of using animal infection in humans, parents were concerned about the invasive procedure, and vaccinated people were often illustrated with a cow’s heads growing from their neck~\cite{Jacobson}. Even though vaccination is an effective way to prevent diseases such as diphtheria, tetanus, pertussis, influenza and measles, almost 1 in 5 children still do not receive routine life-saving immunizations, and an estimated 1.5 million children still die each year of diseases that could be prevented by vaccines that already exist~\cite{who-article}. These fatalities are not only caused by objective reasons such as lack of access to vaccines due to poverty, etc., but also by the unwillingness and fear against vaccines from children's parents. The term “vaccine hesitancy” refers to delay in acceptance or refusal of vaccines despite
availability of vaccine services~\cite{butler2015diagnosing}.
% people being doubtful about vaccinations, or choosing to delay or refuse immunizations even when they are readily available. 
A variety of factors contributes to vaccine hesitancy, including safety concerns, religious reasons, personal beliefs, philosophical reasons, and desire for additional education~\cite{Chephra}. During the COVID-19 pandemic, while the inoculation of  large populations is increasingly important, anti-vaccine narratives are spreading like wildfire endangering public health, human lives, and social order.

With the rise of social media, the dissemination of information (hence, potentially, misinformation) has become easier than ever before. Unsurprisingly, anti-vaccine activists have also began to utilize platforms like Twitter to share their views. As a result, their activism has expanded jurisdictions to include online propaganda. Compared with traditional communication channels, social media offer an unprecedented opportunity to spread anti-vaccination messages and allow communities to form around anti-vaccine sentiment~\cite{Burki}. Social media can amplify the effects of anti-vaccination misinformation: links between susceptibility to misinformation and both vaccine hesitancy and a reduced likelihood to comply with health guidance measures have been shown by multiple studies~\cite{Burki,Broniatowski,Roozenbeek,Johnson}. Based on these findings, vaccine-related misinformation on social media may exacerbate the levels of vaccine hesitancy, creating pockets with low vaccination rates in the United States and globally; this can hamper progress toward vaccine-induced herd immunity, and potentially increase infections related to new COVID-19 variants, possibly leading to vaccine-resistant mutations. For these reasons, understanding vaccine hesitancy through the lens of social media is of paramount importance. Since data access is the first obstacle to attain that, to enable the research community, we built and made public a social media dataset of anti-vaccine content, vaccine misinformation and related conspiracies.

Here we present our dataset that focuses on anti-vaccine narratives on Twitter. The dataset consists of two complementary collections: a) The \textbf{streaming collection} contains tweets collected using Twitter Streaming API from a set of anti-vaccine keywords. The collection started on October 18, 2020  and contains more than 1.8 million tweets posted by 719 thousand unique accounts; b) The \textbf{account collection} contains historical tweets of $\approx 70K$ accounts that engaged in spreading anti-vaccination narratives at some point during 2020, and it contains more than 135 million tweets in total. Additionally, in this paper we present initial statistical analyses of the data, including the frequencies of hashtags, news sources, the accounts' most likely political leaning and geographic locations.

Our IRB-approved dataset only includes tweet IDs of publicly available posts, in compliance with the Twitter Terms of Service. This collection builds on top of the previously published datasets by DeVerna et al.~\cite{DeVerna2021}, that is focused on general vaccine narratives, and complements the previous work by Chen et al.~\cite{Chen} and Lamsal~\cite{Lamsal}, some of the largest Twitter datasets related to COVID-19 discourse published to date. The complete dataset in the form of list of tweet IDs is openly available on GitHub.\footnote{\href{https://github.com/gmuric/avax-tweets-dataset}{https://github.com/gmuric/avax-tweets-dataset}}

\section*{Methods} \label{sec:methods}
\subsection* {Tracked keywords for the streaming collection}
To create a set of keywords that indicate opposition to vaccines, we employed a snowballing sampling technique similar to DeVerna et al.~\cite{DeVerna2021} We started from a small set of manually curated keywords used exclusively in the context of strong vaccine hesitancy that appear on Twitter, such as \#vaccineskill or \#vaccinedamage. Using the Twitter Streaming API and the set of seed keywords, we collected the data for one day (October 18, 2020), after which we extracted other keywords that co-occur with the seed keywords. We added the newly collected keywords to the list of the seed keywords, checking them manually for relevance and excluding irrelevant ones. We then repeated this step several times until we reached about 60 keywords and exhausted the significant co-occurrences. The Twitter API can be queried with a sub-string of a longer keyword and it will return the tweets that contain the sub-string. For example, the keyword ''novaccine'' will return the tweets that contain ''novaccineforme''. We attempted to keep only the most informative and relevant stem words, to capture most vaccine-related tweets and to avoid collecting less relevant ones. 
% For our collection, we used some keywords that overlap, due to dynamic process of keyword curation. 
The list of all keywords used to collect the \textit{streaming collection} is listed in Table~\ref{tab:keywords}.

\begin{table}[H]
  \centering
  \caption{A set of keywords used for our streaming collection}
  \label{tab:keywords}
\scalebox{0.8}{
\begin{tabular}{lllll}
\toprule
Keyword & Tracked since &  & Keyword & Tracked since  \\
\midrule
abolishbigpharma & 12/30/2020 & & noforcedflushots & 12/30/2020\\
antivaccine & 12/30/2020 & & NoForcedVaccines & 10/19/2020\\
ArrestBillGates & 10/19/2020 & & notomandatoryvaccines & 12/30/2020\\
betweenmeandmydoctor & 12/30/2020 & & NoVaccine & 10/19/2020\\
bigpharmafia & 10/19/2020 & & NoVaccineForMe & 10/19/2020\\
bigpharmakills & 12/30/2020 & & novaccinemandates & 12/30/2020\\
BillGatesBioTerrorist & 10/19/2020 & & parentalrights & 12/30/2020\\
billgatesevil & 12/30/2020 & & parentsoverpharma & 12/30/2020\\
BillGatesIsEvil & 10/19/2020 & & saynotovaccines & 12/30/2020\\
billgatesisnotadoctor & 12/23/2020 & & stopmandatoryvaccination & 10/19/2020\\
billgatesvaccine & 12/14/2020 & & syringeslaughter & 12/30/2020\\
cdcfraud & 10/19/2020 & & unvaccinated & 12/30/2020\\
cdctruth & 10/19/2020 & & v4vglobaldemo & 12/30/2020\\
cdcwhistleblower & 10/19/2020 & & vaccinationchoice & 12/30/2020\\
covidvaccineispoison & 12/23/2020 & & VaccineAgenda & 10/19/2020\\
depopulation & 10/19/2020 & & vaccinedamage & 10/19/2020\\
DoctorsSpeakUp & 10/19/2020 & & vaccinefailure & 10/19/2020\\
educateb4uvax & 10/19/2020 & & vaccinefraud & 10/19/2020\\
exposebillgates & 12/30/2020 & & vaccineharm & 10/19/2020\\
forcedvaccines & 12/30/2020 & & vaccineinjuries & 12/30/2020\\
Fuckvaccines & 10/19/2020 & & vaccineinjury & 10/19/2020\\
idonotconsent & 12/30/2020 & & VaccinesAreNotTheAnswer & 10/19/2020\\
informedconsent & 12/14/2020 & & vaccinesarepoison & 10/19/2020\\
learntherisk & 10/19/2020 & & vaccinescause & 10/19/2020\\
medicalfreedom & 12/30/2020 & & vaccineskill & 10/19/2020\\
medicalfreedomofchoice & 12/30/2020 & & vaxxed & 11/02/2020\\
momsofunvaccinatedchildren & 12/30/2020 & & yeht & 11/02/2020\\
mybodymychoice & 12/30/2020 & & &\\

\bottomrule
\end{tabular}
}
\end{table}

\subsection* {Collecting tweets for account collection}
First, we identified a randomly-sampled set of $\approx 70K$ accounts that appeared in the \textit{streaming collection} and were engaged in anti-vaccine rhetoric toward the end of 2020, either by tweeting some of the tracked keywords or by retweeting tweets that contain some of the tracked keywords. Then, for those accounts, we collected their historical tweets, using the Twitter Search API. By leveraging the Twitter's \textit{Academic Research Product Track}, we were able to access the full archival search and overcome the limit of 3,600 historical tweets of the standard API.\footnote{https://developer.twitter.com/en/solutions/academic-research/products-for-researchers}

\subsection* {Calculating the \textit{avax} score}
The \textit{avax} score (short for \textit{anti-vax score}) is a measure we defined and used to quantify an account's vaccine hesitancy level. It is the proportion of an account's tweets that contain any of the anti-vaccine keywords out of all tweets of the given account, and it
is calculated as follows:

\begin{equation}
    X_{i} = \frac{N^{x}_{i}}{N_{i}},
\end{equation}

where, $N^{x}_{i}$ is the number of tweets of account $i$ that contain some of the anti-vaccination keywords from Table~\ref{tab:keywords}, and $N_{i}$ is the number of all tweets of account $i$. Note a distinction between the \textit{retweets} and the \textit{tweets} of an account, where the former are posts of other users that an account shares to most likely endorse or rebroadcast, whereas the latter are original posts created by that account. Therefore, we derive two separate versions of \textit{avax} score, one related to account's retweets $X^{RT}_{i}$, and the second one related to account's original tweets $X^{T}_{i}$.

\subsection* {Calculating accounts' political leanings}
We calculate the accounts' political leaning by measuring the political bias of the media outlets they share. We use a methodology proposed in prior work~\cite{Bovet2019,Badawy2019,Emilio}, and we identify a set of 90 prominent media outlets and accounts that appear on Twitter. Each of these outlets and associated Twitter accounts are placed on a political spectrum (left, lean left, center, lean right, right) per ratings provided by the non-partisan service \textit{allsides.com}\footnote{https://www.allsides.com/unbiased-balanced-news}. For each  account in the dataset, we keep a record of all retweets and the original tweets that contain a domain from the set of selected media outlets. The accounts’ political bias is calculated as the average political bias of all media outlets they shared content from. We again differentiate between original tweets and retweets, and therefore we derive two measures of political leaning.

\subsection* {Identifying low- and high-credibility media sources}
We leverage \textit{urllib}, the Python URL handling module to parse URLs found in the data set. Each URL is broken into several components, including addressing scheme, network location, path etc. A third party data set that contains mis/disinformation domains is used as ground truth to tag the domain names.~\cite{Iffy} For URLs that are not in the data set we query the Media Bias/Fact Check (MBFC) for further identification. Because URL shortening services like Bitly (bit.ly) are widely used on Twitter, such URLs appear with high frequencies. However, they are discarded since they don't refer to proper media sources. Domain names of popular news aggregators and social networks including twitter.com, facebook.com, instagram.com and youtube.com are ignored as well.

\subsection* {Generating geolocation distribution maps}
The number of tweets originating from a geographic region indicates how active accounts are in that region. We adopt a simple approach to infer tweets' geolocations. We use the information on self-reported location of the account and match it to a corresponding US state. To calculate the average activity level per state, the absolute number is normalized by the 2010 Census-reported population of that state as follows:

\begin{equation}
    Avg_{i} = \frac{N_{i}}{P_{i}}\times 1,000,000
\end{equation}

where, $N_{i}$ is the number of tweets of state $i$ and $P_i$ is its 2010 population. Note that we did not generate the geolocation map for the \textit{account collection} as it contains a relatively small number of accounts with self-reported location. 

\subsection* {Topic network analysis}
A topic network is constructed to analyze the co-occurrence of hashtags in the streaming data set. Each node in the graph represents a hashtag, an edge is added if two hashtags occur in the same tweet. The node size is proportional to its degree centrality and the edge weight is the number of times the two hashtags occur. For sake of visualization, nodes with fewer than 25 neighbors are ignored. To investigate the community structure of the network we used the Louvain Algorithm~\cite{Blondel2008} on the topic network, that provides further insights on which anti-vaccine topics are linked together.

\section*{Data overview}
As of this writing (early May 2021), we have collected over 137 million tweets in total. The \textit{streaming collection} focuses on a set of anti-vaccine keywords from Table~\ref{tab:keywords}. The \textit{account collection} on the other hand contains the historical activities of susceptible anti-vaccine accounts, making it a significantly larger data set compared to the streaming one. Although researchers have been collecting data related to COVID-19 vaccines~\cite{DeVerna2021}, as far as we know, there are no public data sets focused specifically on susceptible accounts' historical activities on Twitter. The basic statistics on the two datasets are shown in Table~\ref{tab:stats}. Because the data collection is still ongoing, the statistics shown below are subject to vary in future versions of the data. In the following sections we will analyze \textit{streaming collection} and \textit{account collection} separately.

\begin{table}[H]
  \centering
  \caption{Basic statistics on streaming collection and account collection}
  \label{tab:stats}
\scalebox{0.8}{
\begin{tabular}{lrr}
\toprule
 & Streaming collection & Account collection  \\
\midrule
Number of tweets & 1,832,333 & 135,949,773\\
Number of accounts & 719,652 & 78,954 \\
Average tweet per account & 2.5 & 1,721.8 \\
Verified accounts & 9,032 & 239 \\
Accounts with location & 5,661 & 363 \\
Oldest tweet & 2020-10-19 & 2007-03-06 \\
Most recent tweet & 2021-04-21 & 2021-02-02 \\
\bottomrule
\end{tabular}
}
\end{table}

\section* {Streaming collection}
\begin{figure}[H]
    \centering
    \includegraphics[width=6.8in]{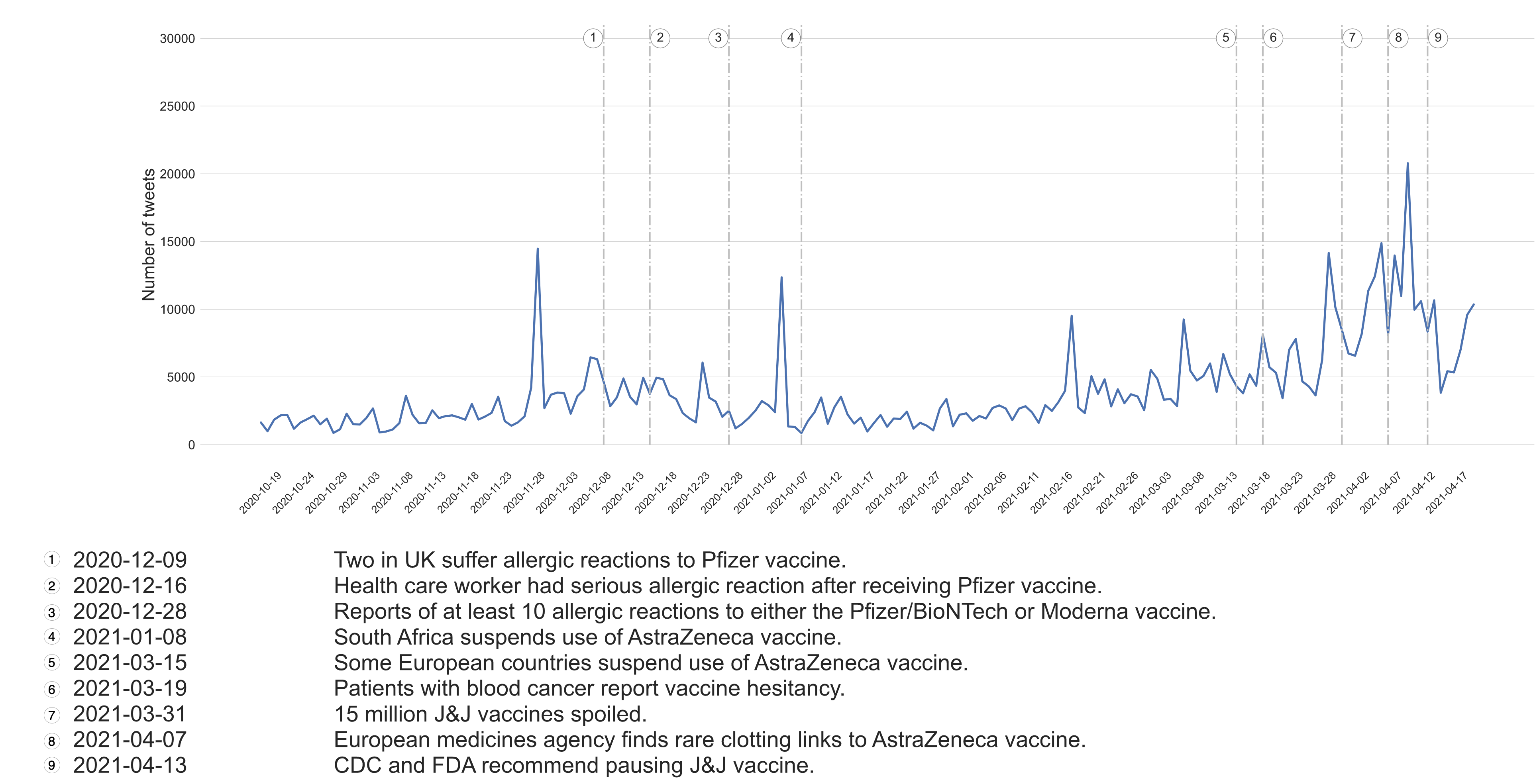}
    \DeclareGraphicsExtensions.
    \caption{Number of tweets over time in the streaming collection. The time of adverse events related to vaccine are marked by dashed vertical lines. Further descriptions about the news are provided in the table below.}
    \label{fig:tweets_date_stream}
\end{figure}

The \textit{streaming collection} consists of 1.8 million tweets by 719K unique accounts, and spans the period from October 18, 2020 to April 21, 2021. As shown in Figure~\ref{fig:tweets_date_stream}, the number of relevant tweets in the \textit{streaming collection} is gradually increasing from the start date. The chatter is relatively stable with small spikes that usually occur on or around the time of major announcements regarding vaccine research or vaccine authorizations. This is expected, as the news usually drive the discussion on Twitter. We observe a large spike in activity near the end of November 2020, that is not caused by any single event but rather by the increased activity of a small number of accounts. 

\begin{figure}[H]
     \centering
     \begin{subfigure}[b]{0.49\textwidth}
         \centering
         \includegraphics[width=\textwidth]{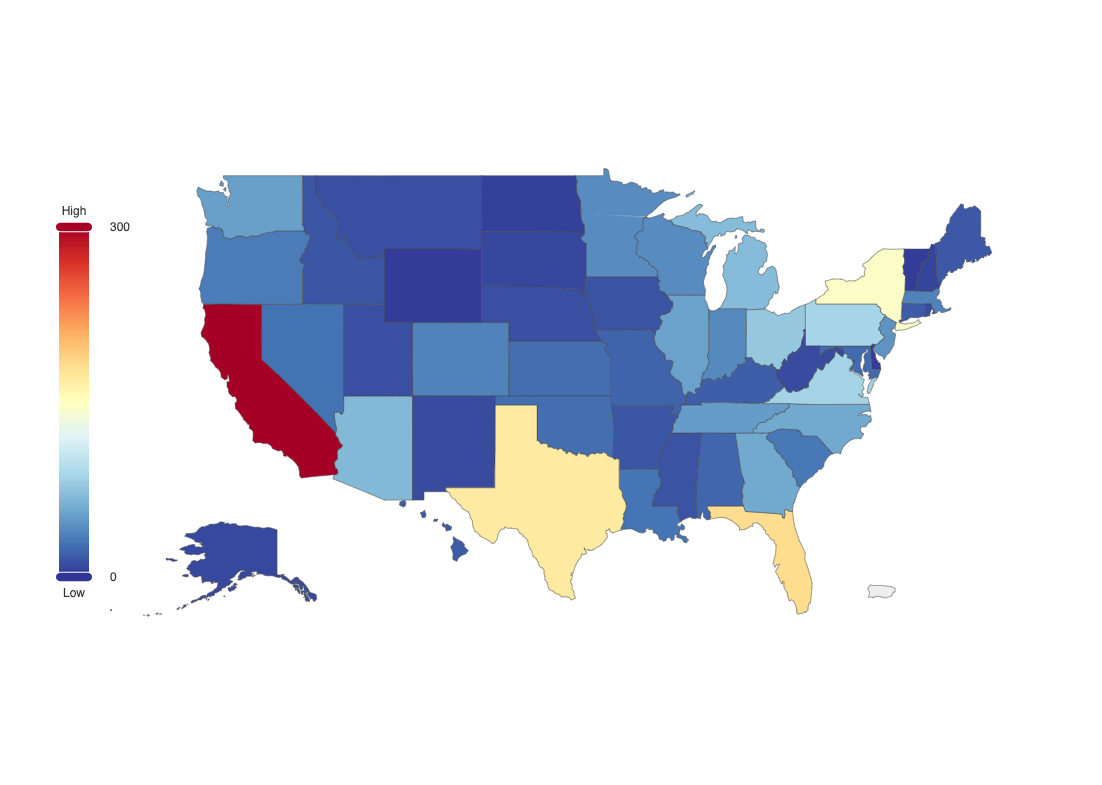}
         \caption{Absolute number of tweets.}
         \label{fig:map-stream}
     \end{subfigure}
     \hfill
     \begin{subfigure}[b]{0.49\textwidth}
         \centering
         \includegraphics[width=\textwidth]{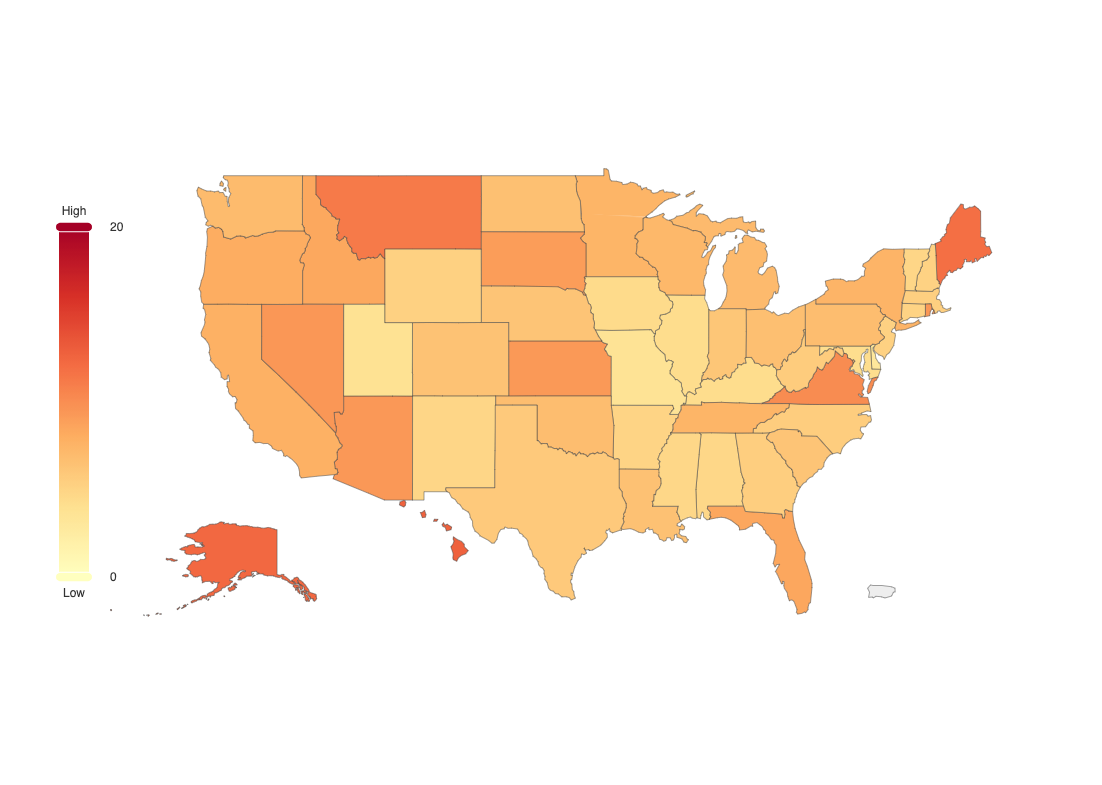}
         \caption{Number of tweets per 1 million people.}
         \label{fig:map-stream-avg}
     \end{subfigure}
        \caption{Geographical distribution of the tweets from the streaming collection originating in the United States: a) absolute numbers and b) normalized by the state population.}
        \label{fig:maps-stream}
\end{figure}

The vast majority of tweets come from countries with predominantly English speaking population. Approximately 68\% of all tweets originated in the United states, 12.5\% in Great Britain, 5.5\% in Canada, 1.2\% in Ireland, 1.1\% in Australia and the rest of the tweets come from other countries. In Figure \ref{fig:maps-stream}, we show the distribution of the tweets' geolocations in the United States. As expected, states with a large population, such as California, Texas, Florida and New York have more tweets in absolute terms (Figure \ref{fig:map-stream}). The number of tweets normalized by state population is provided in Figure \ref{fig:map-stream-avg}, where Hawaii, Alaska, and Maine rank 1st, 2nd, 3rd places, respectively.

Table \ref{tab:top-10-hahstags-stream} lists the top 15 tweeted hashtags in the \textit{streaming collection}. The count is the total number of times a hashtag appears, and the frequency quantifies the proportion of tweets that contain a specific hashtag out of all tweets with a hashtag. Note that there are many tweets without any hashtag and many tweets with a hashtag contain more than one hashtag. Besides the most common general hashtags that are expected to be found, such as ''\#vaccine'' and ''\#covid19'', we observe a high proportion of hashtags that carry strong anti-vaccine sentiment such as ''\#novaccineforme'', ''\#vaxxed'' and ''\#vaccineinjury''. For example, the hashtag ''\#novaccineforme'' can be found in more than 25K tweets, accounting for 6.6\% of all tweets in the \textit{streaming collection} that contain a hashtag. A large set of common hashtags is related to some of the debunked conspiracy theories, claiming that there is a global plot by rich individuals to reduce world population, often expressed through hashtags such as ''\#depopulation'', ''\#billgatesbioterrorist'' and ''\#arrestbillgates''. Another set of very frequent hashtags looks benign on the surface: Hashtags such as ''\#learntherisk'' and ''\#informedconsent'' appear to communicate genuine concerns about the safety of the vaccines; however, those hashtags are usually a decoy and are very often used by the same accounts who are strongly opposing vaccination and who otherwise often use more explicit anti-vaccine hashtags.

\begin{table}[h]
\centering
\scalebox{0.8}{
\begin{tabular}{lrr}
\toprule
Hashtag &  Count &   Frequency (\%) \\
\midrule
vaccine               &  41,069 &  10.66 \\
vaccines              &  33,050 &   8.58 \\
covid19               &  26,616 &   6.91 \\
novaccineforme        &  25,642 &   6.66 \\
learntherisk          &  23,340 &   6.06 \\
billgatesbioterrorist &  20,197 &   5.24 \\
study                 &  20,166 &   5.23 \\
novaccine             &  19,410 &   5.04 \\
mybodymychoice        &  19,166 &   4.97 \\
informedconsent       &  16,578 &   4.30 \\
depopulation          &  15,021 &   3.90 \\
vaxxed                &  12,691 &   3.29 \\
vaccineinjury         &  12,640 &   3.28 \\
vaccination           &  10,873 &   2.82 \\
arrestbillgates       &   9,991 &   2.59 \\
\bottomrule

\end{tabular}
}
\caption{Top 15 hashtags in streaming data set. The count is the total number of times a hashtag appears. The frequency quantifies the proportion of tweets that contain a specific hashtag out of all tweets with a hashtag.}
\label{tab:top-10-hahstags-stream}
\end{table}

\begin{figure}[H]
\centering
\includegraphics[width=4in]{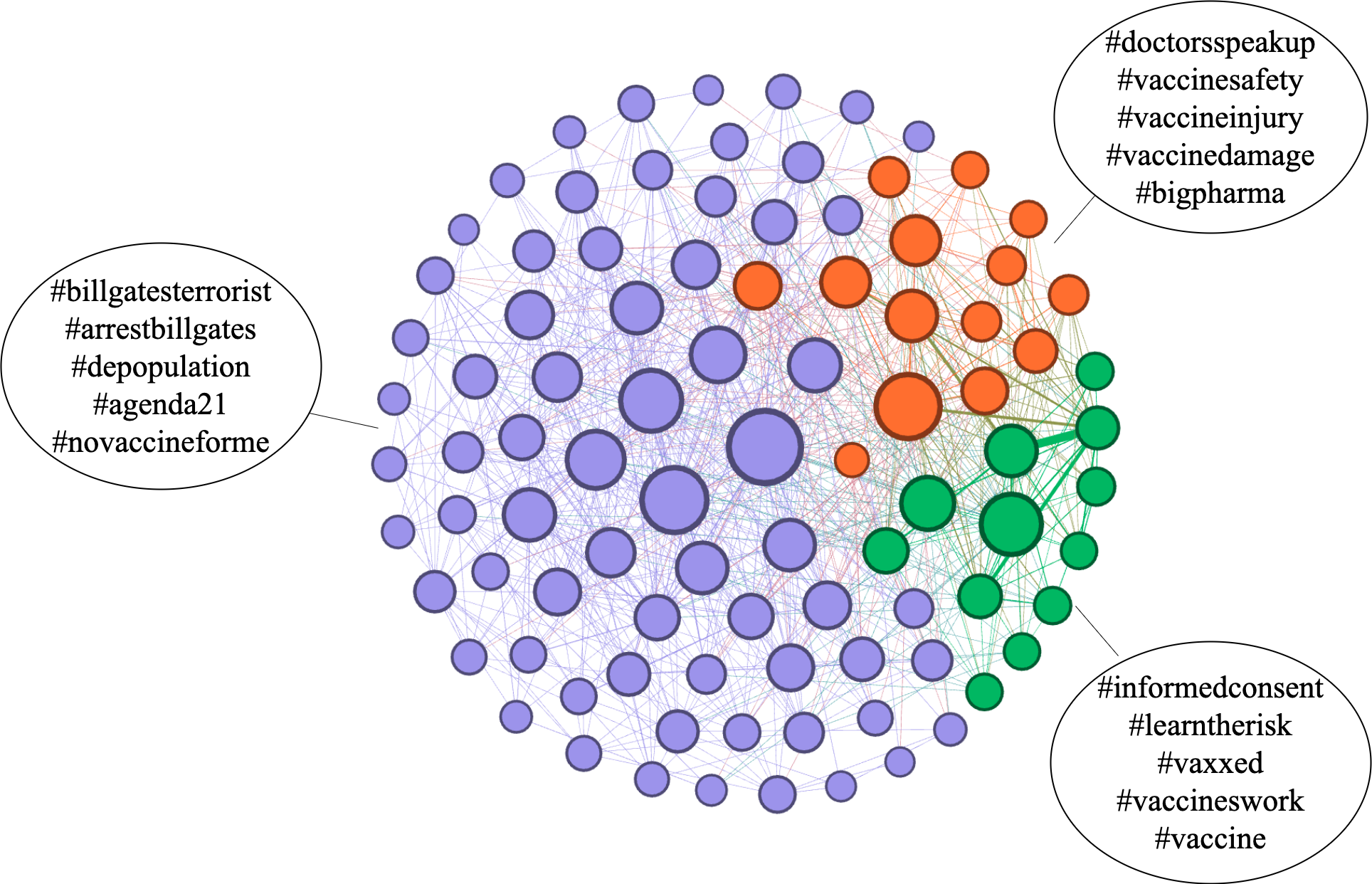}
\DeclareGraphicsExtensions.
\caption{An overview of the prominent hashtags in the data set, clustered into 3 communities. For readability we didn't show all the node labels.}
\label{fig:topic_network}
\end{figure}

To get further insights into co-appearing anti-vaccine narratives, we explore which anti-vaccine topics usually co-occur together. We run the Louvain community detection algorithm on the topic co-occurrence network as described in the Methods section. The topic network is illustrated in Figure \ref{fig:topic_network}. We identify three distinct communities, and all of them contain anti-vaccine keywords, but with different focus on topics. The largest topic community (in purple) is focused on debunked claims around the conspiracy narrative that the vaccine is a plot of rich people to reduce world population. The second topic community (orange) mostly focuses on vaccine safety as hashtags such as ''\#doctorsspeakup'', ''\#vaccinesafety'' and ''\#vaccineinjury''' appear very often. The smallest topic community (green) contains a mixture of various hashtags that range from strongly anti-vaccine such as ''\#informedconsent'', ''\#learntherisk'', ''\#vaxxed'', some neutral hashtags such as ''\#vaccine'', to some pro-vaccine hashtags such as ''\#vaccineswork''.

\begin{figure}[H]
    \centering
    \includegraphics[trim={0.3in 0 0 0},clip,width=4.2in]{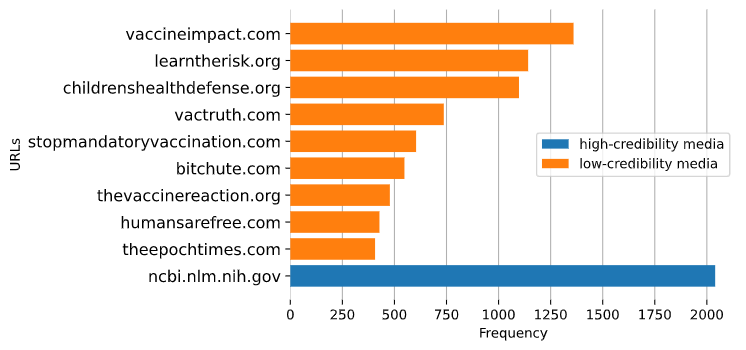}
    \DeclareGraphicsExtensions.
    \caption{Top 10 news sources in the streaming collection. The URL's of the news aggregators and the large social platforms are omitted.}
    \label{fig:url_freq_stream}
\end{figure}

Vaccine hesitancy is usually fueled by misinformation originating in Websites with questionable credibility. In Figure ~\ref{fig:url_freq_stream}, we illustrate the top 10 URLs of various Websites that can be found in the \textit{streaming collection}. The vast majority of those Websites can be found in the \textit{Iffy+ database of low credibility sites}.~\cite{Iffy} Notice that the most commonly shared URL is the Website of the \textit{National Center for Biotechnology Information}\footnote{https://www.ncbi.nlm.nih.gov/} that is part of the \textit{United States National Library of Medicine} (NLM), a branch of the \textit{National Institutes of Health} (NIH). NCBI houses PubMed, the largest bibliographic database for biomedical literature. This can give a false impression that the tweets from the \textit{streaming collection} share information from legitimate scientific sources. However, most of the papers from PubMed are cited with false and misleading conclusions. Sometimes, anti-vaccine advocates would share legitimate scientific papers documenting rare side-effects of the vaccines, while over-emphasizing the observed adverse effects and calling for vaccine boycotts. Sharing a scientific study in a tweet provides an illusion of credibility. Cherry-picking desirable sentences and relying on the fact that the majority of the audience will not make an effort to read a scientific paper in details, makes it a very effective strategy for manipulation.

\section* {Account collection}
\begin{figure}[h]
     \centering
     \begin{subfigure}[b]{0.41\textwidth}
         \centering
         \includegraphics[width=\textwidth]{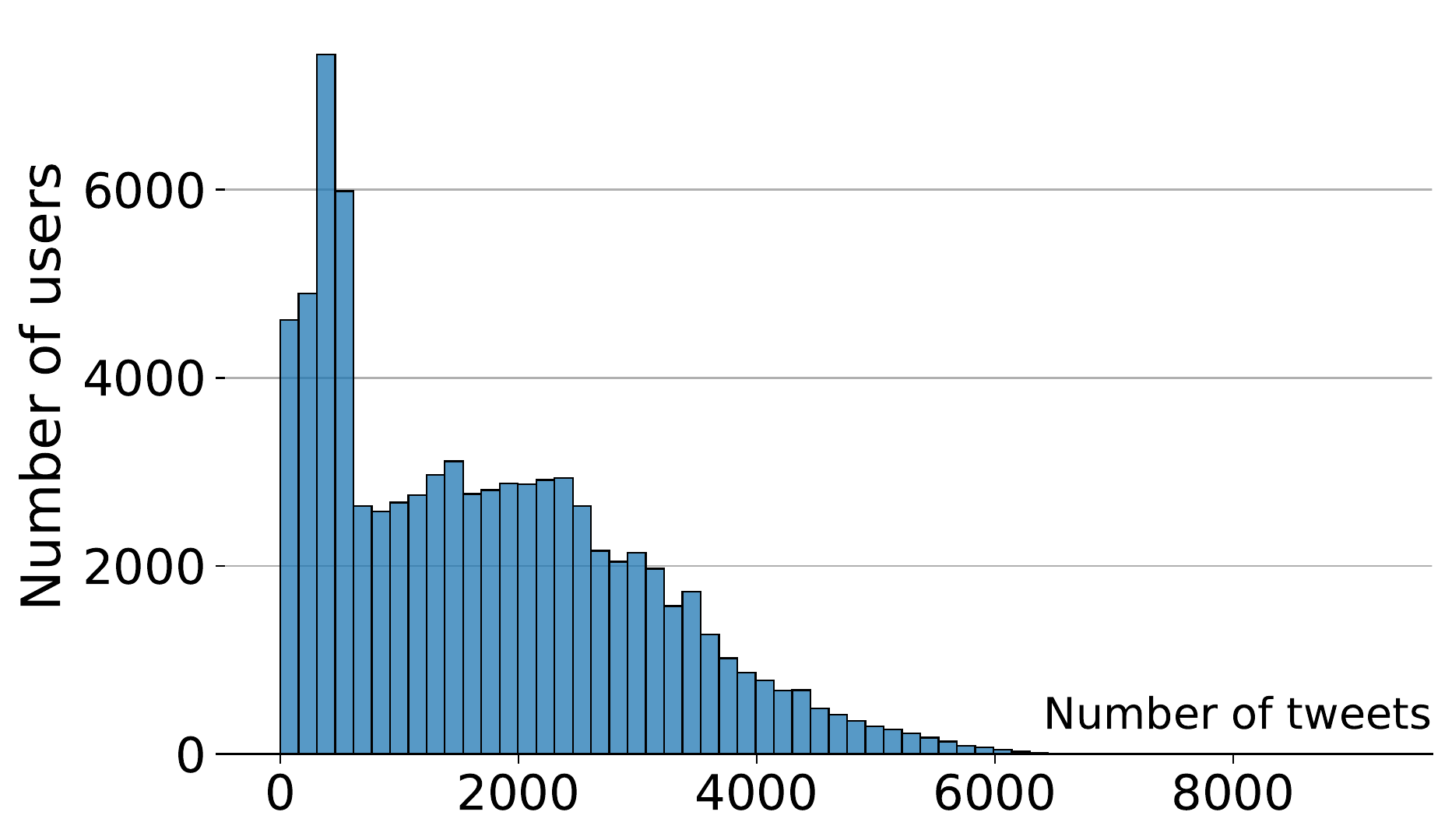}
         \caption{}
         \label{fig:tweets_per_user}
     \end{subfigure}
    \hspace{.3in}
     \begin{subfigure}[b]{0.41\textwidth}
         \centering
         \includegraphics[width=\textwidth]{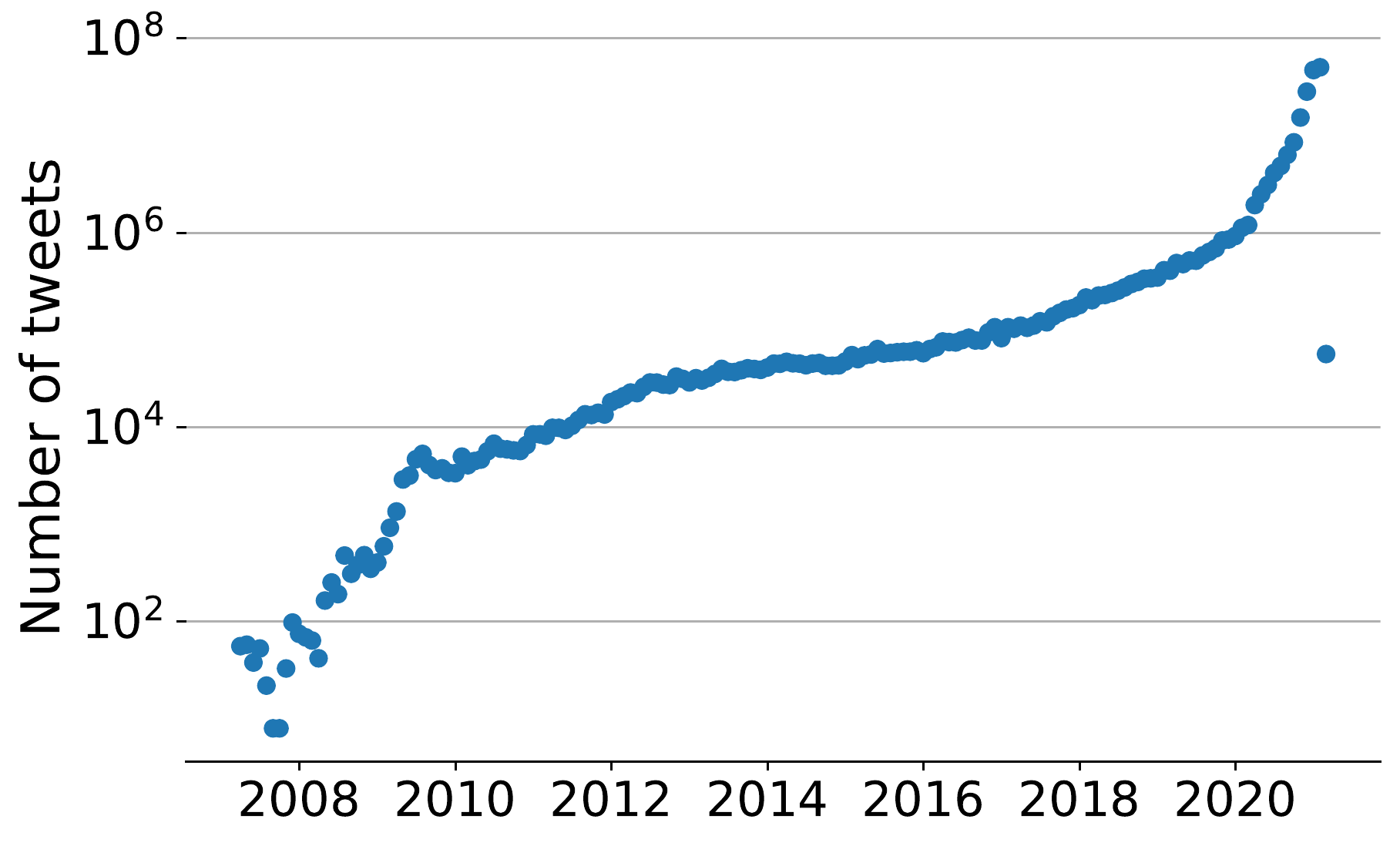}
         \caption{}
         \label{fig:tweets_per_year}
     \end{subfigure}
     
        \caption{Account collection: a) Distribution of tweets per account; b) Distribution of tweets over time}
        \label{fig:user_tweets_stats}
\end{figure}

The \textit{account collection} differs from the \textit{streaming collection} as it is focused on historical tweets from a set of accounts. The process of collecting historical tweets is explained more in details in the Methods section. The current account collection consists of more than 135 million tweets published by over 78K unique accounts, and spans the period from March 3, 2007 to February 8, 2021. In Figure~\ref{fig:user_tweets_stats}, we illustrate some of the most important statistics from this data collection. Figure~\ref{fig:tweets_per_user} shows a distribution of tweets per account. Approximately half of the accounts published fewer than 1,500 tweets, 40\% of the accounts have more than 2,000 tweets and 1.5\% have more than 5,000 tweets. Figure~\ref{fig:tweets_per_year} shows the number of tweets over time. Most of the tweets originate in the year 2020, with the oldest tweet dating back to 2007. For $\approx 70\%$ of the accounts, the oldest collected tweet dates from 2020. There is a significant portion of accounts whose historical tweets date much earlier with $\approx 18\%$ of accounts with their earliest tweet dating before 2018 and $\approx 6.8\%$ accounts with the tweets before 2014. Such relatively long-spanning collection of historical tweets at the account level may allow for a comprehensive temporal analysis of how vaccine hesitancy developed on Twitter over the years. 

The most common 15 hashtags appearing in the \textit{account collection} are displayed in Table~\ref{tab:top-10-hahstags-user}. Beside the common covid-related hashtags, there are many hashtags referring to US politics. During the period around the US 2020 presidential election, the accounts that appear in our collection were particularly active. Hence, we can see many politically-motivated narratives in the data originating during that time period. Accounts that share common misinformation related to the vaccines are prone to conspiratorial thinking and very often shares other conspiracy narratives, usually the politically charged ones. It is known that the population susceptible to such narratives strongly skews conservative~\cite{Emilio}, therefore we expected that a large number of accounts in the \textit{account collection} is right leaning.

\begin{table}[h]
\centering
\scalebox{0.8}{
\begin{tabular}{lrr}
\toprule
Hashtag &  Count &   Frequency (\%) \\
\midrule
covid19      &  474,481 &  2.55 \\
endsars      &  203,297 &  1.09 \\
maga         &  164,332 &  0.88 \\
coronavirus  &  158,574 &  0.85 \\
trump        &  156,262 &  0.84 \\
stopthesteal &  121,069 &  0.65 \\
trump2020    &  115,002 &  0.62 \\
breaking     &  111,274 &  0.60 \\
obamagate    &  110,046 &  0.59 \\
covid        &  106,095 &  0.57 \\
china        &   98,026 &  0.53 \\
oann         &   96,943 &  0.52 \\
antifa       &   79,157 &  0.43 \\
biden        &   77,728 &  0.42 \\
fakenews     &   66,599 &  0.36 \\
\bottomrule

\end{tabular}
}
\caption{Top 15 hashtags in \textit{account collection}. The count is the total number of times a hashtag appears. The frequency quantifies the proportion of tweets that contain a specific hashtag out of all tweets with a hashtag.}
\label{tab:top-10-hahstags-user}
\end{table}

\begin{figure}[H]
     \centering
     \begin{subfigure}[b]{0.45\textwidth}
         \centering
         \includegraphics[trim={0 0 0 0.45in},clip,width=\textwidth]{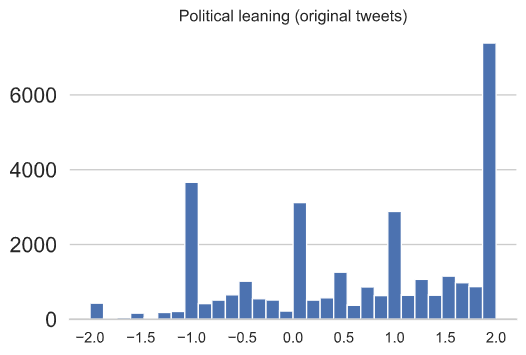}
         \caption{}
         \label{fig:orig-distribution}
     \end{subfigure}
     \hfill
     \begin{subfigure}[b]{0.45\textwidth}
         \centering
         \includegraphics[trim={0 0 0 0.45in},clip,width=\textwidth]{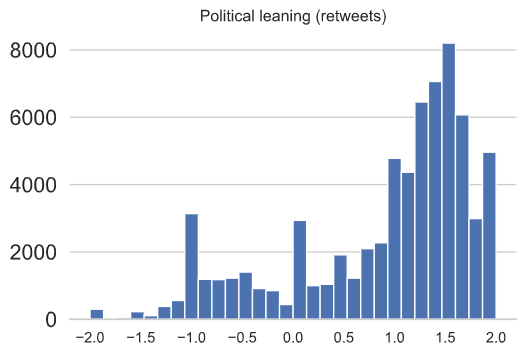}
         \caption{}
         \label{fig:rtw-distribution}
     \end{subfigure}
        \caption{Distribution of accounts' most likely political leanings from the \textit{account collection}: a) Calculated from the original tweets; b) Calculated from the retweets}
        \label{fig:three graphs}
\end{figure}

In Figure~\ref{fig:three graphs}, we show the distribution of the accounts' political leanings. An estimation of the political leaning of an account is based on their media diet (see Methods). The political leaning of the accounts inferred from their \textit{original tweets} is illustrated in Figure~\ref{fig:orig-distribution}; the political leaning of the accounts inferred from their \textit{retweets} is illustrated in Figure~\ref{fig:rtw-distribution}. In both cases, the distribution of the most likely political orientation skews strongly to the right. Such results are not surprising as they align to earlier studies showing that  political orientation is a strong predictor of vaccine hesitancy in the US.~\cite{Fridman2021,Ruiz2021}

\begin{figure}[H]
    \centering
    \includegraphics[trim={0.3in 0 0 0.45in},clip,width=4.2in]{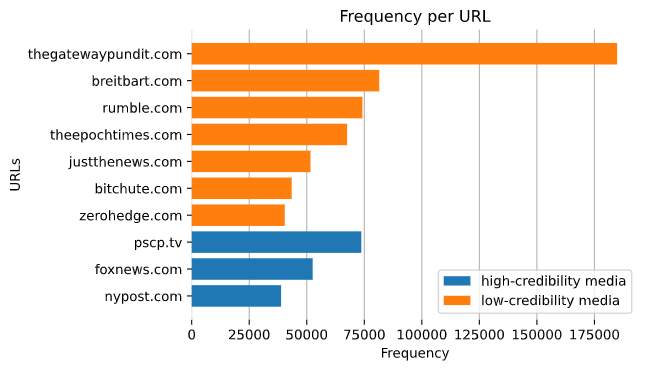}
    \DeclareGraphicsExtensions.
    \caption{Top 10 URLs in \textit{account collection}.}
    \label{fig_url_freq}
\end{figure}

Finally, Figure \ref{fig_url_freq} shows that many far-right news media sites appear frequently in the \textit{account collection}. The Gateway Pundit\footnote{www.thegatewaypundit.com}, which is known for publishing falsehoods, hoaxes, and conspiracy theories, occurs more than 180,000 times. Other far-right media such as Breitbart News\footnote{www.breitbart.com} and the Epoch Times\footnote{www.theepochtimes.com} also appear very often. Among the high-credibility URLs, Periscope\footnote{www.pscp.tv}, a live video streaming service acquired by Twitter in January 2015 but no longer active since March 21, 2021, tops the list.~\cite{twitter-preiscope-faq} While it is considered as the mainstream service for sharing videos, many susceptible accounts misuse this service to spread dis/misinformation, ranging from political conspiracies such as ''QAnon'' and ''Obamagate'' to various anti-vaccine narratives. For example, a video titled ''\textit{Dr.SHIVA LIVE: Truth Freedom \& Health. Vs. Fear Fauci \& Fascism.}'', published by a self-proclaimed ''doctor'' (the account has been since suspended by Twitter), had more than 62K viewers. Considering other sources that usually fall in the group of mainstream news media sites such as Fox News\footnote{www.foxnews.com} and New York Post\footnote{www.nypost.com}, conspiracy spreaders selectively quote reports from these sources to increase the credibility of their false claims.

\section* {Discussion}

In this paper, we presented a comprehensive data set consisting of the tweets related to  anti-vaccination narratives. The dataset contains two collections: \textit{a) Streaming collection}, which leverages the Twitter Streaming API and contains the tweets with one or more tracked keywords; \textit{b) Account collection}, which is a longitudinal data set, tracking the historical tweets of a sample of accounts who got engaged with anti-vaccination narratives. We characterized the data in several ways, including frequencies of the prominent keywords, news sources, geographical location of the accounts and their political leaning. 

Although the data sets give an overview of vaccine hesitancy on Twitter, potential limitations warrant some considerations. First, our \textit{streaming collection} relies on a defined set of keywords. The anti-vaccine lingo is constantly evolving as the COVID-19 pandemic unfolds. While we have made our best efforts to find most representative keywords, they might not fully cover all anti-vaccine topics. The set of keywords we used is designed to capture the strongest anti-vaccine sentiments, and can miss the various nuances in the multifaceted nature of vaccine hesitancy. Secondly, this dataset should not be used to draw conclusions for the general population, as the Twitter user population is younger and more politically engaged than the general public~\cite{twitter-user-profile}, which means our data may be biased in various ways. Finally, to prevent COVID-19 misleading information to spread, Twitter has enacted specific rules and policies\footnote{https://help.twitter.com/en/rules-and-policies/medical-misinformation-policy}. The accounts violating these rules and policies may be banned by Twitter, making their tweets unreachable. 

Besides the \textit{streaming collection} that tracks the tweets as they appear in real-time, the second main contribution of this paper is the \textit{account collection} that can be used to provide further insights into the accounts that engage in anti-vaccine narratives. In future work, we intend to track the longitudinal characteristics of accounts engaging with anti-vaccine narratives, to gain better insights into the socioeconomic, political, and cultural determinants of vaccine hesitancy. Our intention by publishing this paper and datasets is to provide researchers with the assets to enable further exploration of issue revolving around vaccine hesitancy and study them through the lens of social media.

\section*{Usage Notes}
The dataset is released in compliance with the Twitter’s Terms \& Conditions and the Developer’s Agreement and Policies. Researchers who wish to use this dataset must agree to abide by the stipulations stated in the associated license and conform
to Twitter’s policies and regulations.

\section*{Data availability}
The data is available at: \href{https://github.com/gmuric/avax-tweets-dataset}{https://github.com/gmuric/avax-tweets-dataset}

\section*{Acknowledgements}

The authors appreciate the support of the Annenberg Foundation.

\section*{Competing interests}
The authors declare no competing interests.

\bibliography{sample}

\end{document}